\documentclass[12pt]{article}

\newcommand{\atg}{{\rm\, arctg\, }}

\newcommand{\la}{{\rm \lambda }}

\newcommand{\wpar}{{w_{_\parallel} }}
\newcommand{\paral}{{\parallel}}
\newcommand{\wper}{{w_{_\perp} }}

\makeatletter

\renewcommand{\@biblabel}[1]{#1. \hfill}

\input epsf

\begin{document}

\begin{center}
{\bf \Large
Einstein-Rosen Bridges and the Characteristic Properties\\
of Gravitational Lensing by Them.}
\end{center}

\begin{center}
A.A. Shatskiy
\end{center}

\begin{center}
Astro Space Center, Lebedev Physical Institute, Russian Academy of Sciences\\
Profsoyuznaya ul., 84/32, Moscow, 117810, Russia,\\
Received November 5, 2004; in final form, January 9, 2004.
\end{center}

{\bf \Large  Abstract }

It is shown that Einstein-Rosen bridges (wormholes) hypothetical
objects that topologically connect separate locations in the
Universe can be static solutions of the Einstein equations. The
corresponding equations for bridges are reduced to a form
convenient for their analysis and numerical solution. The matter
forming the bridge must have a sufficiently hard and anisotropic
equation of state. Our results are compared with a previously
known analytic solution for a bridge, which is a special case of
the general solution in the framework of general relativity. The
deflection of photons by the bridge (gravitational lensing) is
studied.

\section{INTRODUCTION}

In recent years, there have been an increasing number of
publications in relativistic astrophysics devoted to so-called
"wormholes". Another term for these objects proposed by Einstein
and Rosen [1] in 1935 is "bridge".

There are several different definitions of bridges, depending on
the presence or absence of event horizons in them. A common
feature of all these definitions is that the bridge connects two
asymptotically at spatial regions. The location of such a
connection is the bridge, and its central part is called the
bridge's throat (Fig.1). Space-time is strongly curved near the
throat. We shall consider here only traversable Lorentzian
wormholes, i.e. those through which physical bodies can pass.
Consequently there must be no event horizons in such bridges.

As was proved long time ago [2], bridges can be constructed in
general relativity only from matter with exotic equations of state
(for more details, see below). Such matter has not been found in
the Universe thus far. Therefore, bridge solutions were first
sought in alternative theories of gravity such as the Brans-Dicke
theory and theories involving quantum-gravitational effects.

\begin{figure}[t]
\centering \epsfbox[100 280 500 600]{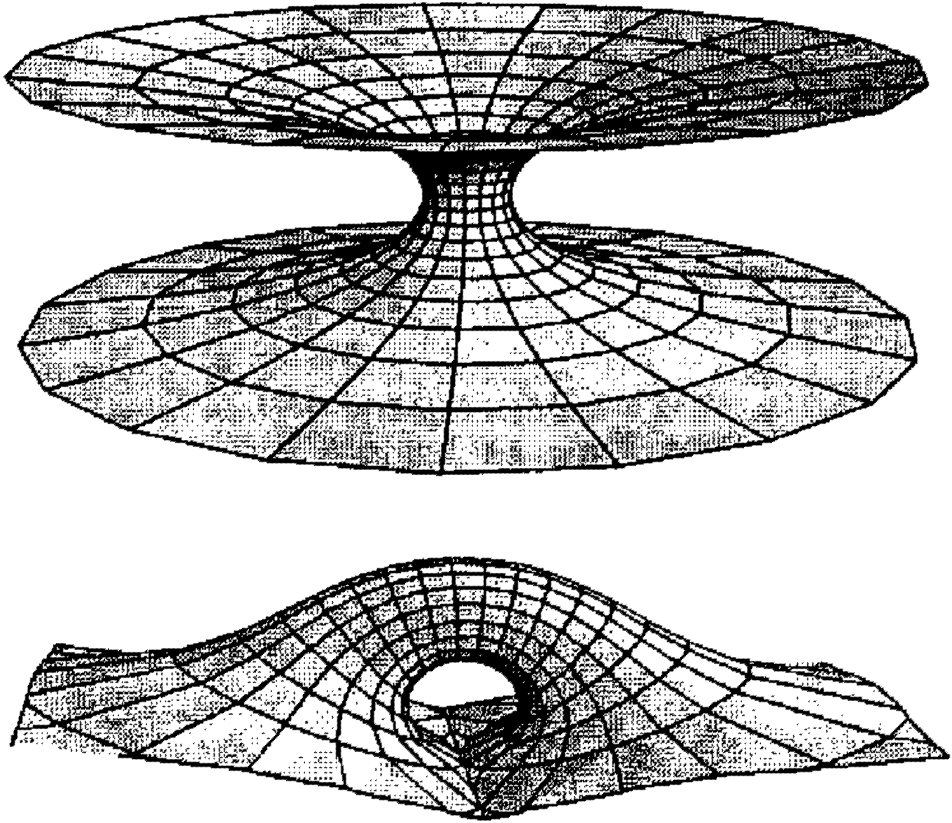} \caption{Spherically
symmetric bridge connecting two parts of the Universe (top panel)
and a non-spherically symmetric "door-handle" bridge connecting
two distant parts of the same Universe (bottom panel).}
 \label{R1}
 \end{figure}

Despite all these problems, bridges have been attracting
increasing interest. This is due in our view to the following four
factors.

(1) The existence of dark matter which may obey an unusual
equation of state.

(2) Bridges could be real and observable astrophysical objects,
such as black holes, whose existence was likewise long rejected by
many scientists.

(3) Huge progress in computer modeling of multidimensional
structures has made it possible to solve numerically many problems
in relativistic astrophysics. Most of these problems cannot be
solved using analytic methods because of their extreme complexity.

(4) There have been serious discussions of solutions of the
equations of general relativity containing time loops, which imply
the existence of time machines (see, for example, [3]).

We shall not discuss here speculations based on

the fourth factor. Our aim is to present the basic ideas of the
physics of bridges and their direct relationship to the theory of
gravity and black-hole physics. The bridges will be considered in
a general-relativitistic treatment.

\section{SPHERICALLY SYMMETRIC SYSTEMS}

Analytic studies can be carried out most easily for spherically
symmetric systems, for which all quantities depend only on time
and radius. The form of the solutions of the equations of general
relativity does not depend only on the symmetry of the system.
Nevertheless, even in cases with the simplest central symmetry
exact solutions have been obtained for only a few cases. These are
the case of dust matter (the pressure is equal to zero) and of the
vacuum condensate (the pressure is the negative of the energy
density). It is assumed in inflationary cosmological models that
this latter equation of state was dominant for matter in the
initial stage of development of the Universe after the Big Bang (a
de Sitter universe) [4].

In more realistic cases, for example, when the pressure is
proportional to the energy density with a positive coefficient (in
particular an ideal photon gas is described by the equation of
state with a coefficient of $1/3$ ), we have no exact solutions,
and only numerical solutions are available. Moreover even for the
simplest models in which all quantities depend only on radius,
exact solutions have been found for only a few cases.

An illustrative example of static solutions are ordinary stars and
planets. However in reality they obey a fairly sophisticated
nonlinear equation of state. As a result, only numerical solutions
can be obtained for these objects in a general-relativitistic
treatment. The spherically symmetric metric in general relativity
can be written:
\begin{equation}
ds^2 = \exp (\nu)\, dt^2 - \exp(\la)\, dr^2 - r^2\, d\Omega^2
\label{0-6}\end{equation} Here, $\nu =2\varphi$ is twice the red
(or violet) shift, $\varphi =(\Delta f )/f$, $f$ is the photon
frequency at infinity near the source, and $d\Omega^2$ is an
element of solid angle. In the general case, the energy momentum
tensor in commoving matter with a spherically symmetric
distribution has the form \footnote{We use the theoretical system
of units in which the speed of light is c =1 and the gravitational
constant is G =1 .}:
\begin{equation}
4\pi T^0_0=\varepsilon \, ,\quad 4\pi T^1_1=-P_{\parallel} \, ,
\quad 4\pi T_2^2=4\pi T_3^3=-P_{\perp} \, .
\label{0-1}\end{equation} It is convenient to describe motion in
the spherically symmetric field using the function $V^2$ which is
the square of the invariant velocity of the particle with respect
to the spheres $r =const$:
\begin{equation}
V^2 = \left({dr\over dt}\right)^2\, e^{\la - \nu} ,\, .
\label{0-5}\end{equation} A more detailed definition of $V^2$ is
presented in [5]. When a falling particle reaches an event horizon
the velocity $V$ reaches unity. The corresponding element of the
interval $ds$ is zero so that $V^2$ is an invariant at the
horizon; i.e. it does not depend on the choice of coordinate
system. In the case of free fall (along geodetic trajectories),
$V^2 =1-\exp (\nu)\cdot const$. In the nonrelativistic limit, the
definition of $V^2$ coincides with the usual definition of the
velocity.

For the collapse of dust matter or a Schwarzschild black hole
$(P_\paral = P_\perp =0)$, we obtain
\begin{equation}
\begin{array}{lll}
ds^2 = (1-r_g/r)^{+1}\, dt^2 - (1-r_g/r)^{-1}\, dr^2 - r^2\, d\Omega^2\, ,\\
\\
V^2 = \left[ (R_0/r) - 1 \right] / \left[ (R_0/r_g) - 1 \right] \,
.
\end{array}
\label{0-7}\end{equation} Here and below $R_0$ is the initial
value of the $r$ coordinate for the falling particle and the event
horizon is defined by $V =1$ .

For an inflating de Sitter universe $(P_\paral = P_\perp
=-\varepsilon )$, we obtain
\begin{equation}
\begin{array}{lll}
ds^2 = (1-r^2/a^2_{_H})^{+1}\, dt^2 - (1-r^2/a^2_{_H})^{-1}\, dr^2
- r^2\,
d\Omega^2\, ,\\
\\
V^2 = \left[ (r/R_0)^2 - 1 \right] / \left[ (a_{_H}/R_0)^2 - 1
\right] \, .
\end{array}
\label{0-8}\end{equation} Here, $H\equiv 1/a$, $H$ is the Hubble
constant.

\section{BRIDGES}

The relation between the circumference of a circle and its radius
is violated in curved space. Therefore, to avoid misunderstanding,
let us introduce the following new coordinates. Let us choose some
point in space as a central point. Next we draw rays from this
point in all directions. Distance along the rays will be measured
by the coordinate $l$ and distance in the perpendicular direction
by the coordinate $r$ . Therefore, the circumference will be (by
definition) $2\pi r$ while the distance from the center of the
circle to its boundary will be (by definition) $l$. In such
coordinates, expression (1) for a static case takes the form

\begin{equation}
ds^2 = \exp (2\phi)\, dt^2 -  dl^2 - r^2\, d\Omega^2
\label{0-6-1}\end{equation}

\begin{figure}[t]

\centering \epsfbox[100 280 500 500]{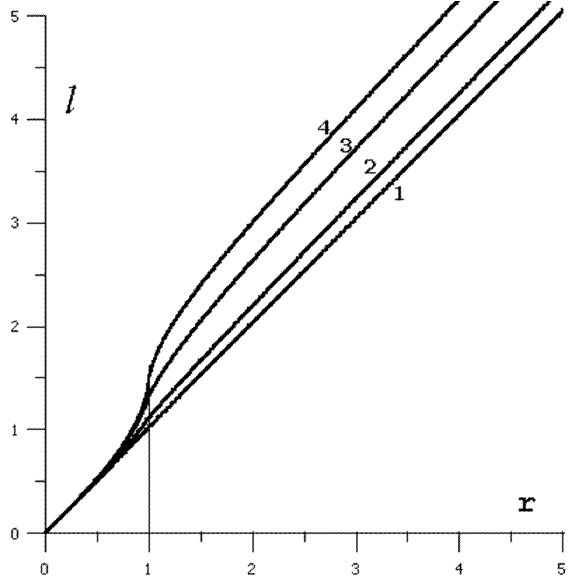}
 \caption{The dependence of $l$ on $r$ for various values of the
Newtonian potential $\varphi_g$ at the surface of a body $(r=1)$:
(1) $\varphi_g =0.1$ nearly coincident with the straight line $l
=r$
 corresponding to $\varphi_g =0$, (2) $\varphi_g =0.5$,
(3) $\varphi_g =0.9$ and (4) $\varphi_g =0.99$.}
 \label{R2}
 \end{figure}

In a curved space $r\ne l$. To illustrate this fact let us put a
small mass (for example, a planet such as the Earth) at the center
of the system. Here, "small mass" implies a body whose Newtonian
gravitational potential at its surface is much less than the
square of the speed of light. For the Earth, this potential is
$\varphi_g\sim 10^{-9}$; for the Sun $\varphi_g\sim 10^{-5}$ and
for a pulsar $\varphi_g\sim 0.5$. The dependence of $l$ on $r$ is
illustrated by the plots in Fig.2. This curving the space-time by
gravitating matter was predicted by Einstein and subsequently
confirmed by numerous experiments.

As the central body becomes heavier it curves space more strongly
and stretches it along the radii. When a certain limit is achieved
the space is curved so strongly that a black hole is formed. What
are the limiting objects just before the formation of black holes,
and are black holes always formed?

One well-known limit is a neutron star (pulsar), whose mass is of
the order of a solar mass and whose radius is about ten
kilometers; the average density of matter in the neutron star is
approximately equal to the density of an atomic nucleus $\sim
10^{13} g/cm^3$.

The above-mentioned curvature of the space is negligible for
normal stars and planets, and is important only for neutron stars
(see curve 2 in Fig.2). The limiting curvature occurs when the
inclination of the curve $l (r)$ with respect to the $r$ axis
increases from $45^\circ$ (the uncurved geometry) to $90^\circ$
(the most curved geometry). In this case, we obtain a zero
increment in $r$ for a nonzero increment in $l$. In Fig.2 curve 4
with $\varphi_g = 0.99$ is closest to this situation (as follows
from calculations this limit is not actually reached: the star
collapses before this occurs when $\varphi_g\sim 0.5$). After the
collapse of a star to a black hole, $l$ becomes imaginary after it
passes the singularity indicated above. This is not a physical but
a coordinate singularity at the black-hole event horizon, which
can be avoided by an appropriate coordinate transformation. The
condition $V =1$ can serve as the invariant definition of the
horizon [see (3)].

There is also an alternative to the curving of physical space by a
black hole. Above a decrease in the $l$ was accompanied by a
decrease in $r$. If there is a black hole at the center of the
system, the coordinate $l$ becomes imaginary at its event horizon.
However there are static solutions in general relativity in which,
after passage of the singularity $r$ again begins to increase as
$l$ is decreased. Such solutions can correspond to bridges.

Let us present here the definition of a bridge that has been
adopted in a number of studies (for example, [6]):

(1) A bridge must have a region where $r$ reaches a local minimum,
$r_0$. This region is called the throat of the bridge, and $r_0$
the throat radius (Fig.1). This is the main difference between a
bridge and any other body.

(2) There must be no event horizons in the bridge; i.e., bodies
falling into the bridge must not reach the speed of light $(V<1)$
and consequently they should be observable for an external
observer.

(3) Bodies can pass through the bridge throat in both directions.

Geometry can be different on the opposite sides of the bridge
throat. We shall consider from here on only spherically symmetric
static solutions for bridges obeying the equation of state
\begin{equation}
 P_{\parallel}=\wpar\varepsilon\, ,
\quad P_{\perp}=\wper\varepsilon\, . \label{0-2}\end{equation}
Here, $w_\paral$ and $w_\perp$ are constant parameters, and the
energy density must not be negative. Such matter is, of course,
idealized, but it enables us to study analytically many important
features of bridges.

The matter forming such bridges should have exotic properties,
since the allowed range of $w_\paral$ and $w_\perp$ is restricted
by the inequality (see Appendix 8.1)
\begin{equation}
\varepsilon > 0\, , \qquad -2\wper < \wpar < -1 \, .
\label{1-14-5}\end{equation}

\begin{figure}[t]
\centering \epsfbox[100 280 500 500]{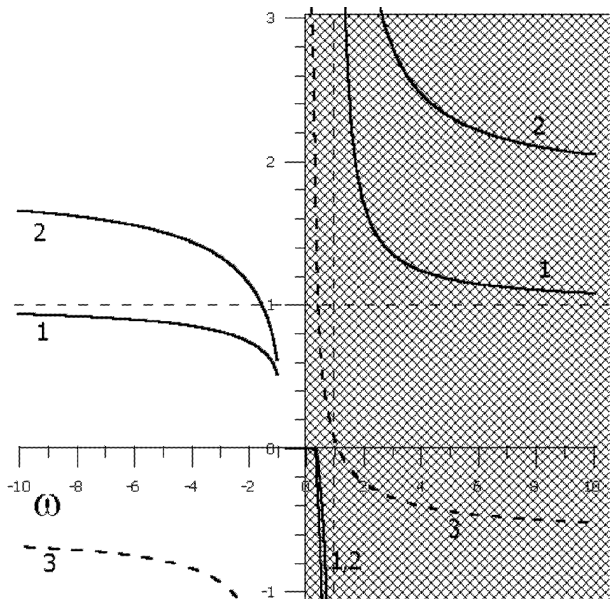}
 \caption{Dependence of the following quantities for bridges
on the equation of state (namely on $w =w_\paral = -w_\perp$): (1)
$r_0 /r_g$, (2) $r_l /r_g$, (3) $\varphi_{max}$. The left-hand
(unshaded) side of the plots corresponds to the physical region
$\varepsilon >0$ and the right-hand (shaded) side to the
unphysical region $\varepsilon <0$.}
 \label{R3}
 \end{figure}

We can see that, in order for a bridge to exist, the signs of
$w_\paral$ and $w_\perp$ must be opposite. Therefore, the equation
of state must be substantially hard and anisotropic.

The mass $m$ and gravitational radius $r_g$ of a bridge are
defined in the same way as for any other body via the field
asymptotics at in infinity:
\begin{equation}
\lim\limits_{r\to\infty}\exp (-\lambda)=1-r_g/r \, ,\quad m =
r_g/2 \, . \label{1-14-6}\end{equation} In addition to its mass, a
bridge is characterized by two dimensions: its length and width.
The bridge width can be identified with the size of its throat
$r_0$. The bridge length can be reasonably related to the physical
characteristics of a body passing through the bridge, for example,
the change in the red (or violet) shift of the signal received
from this body. Therefore, let us define the bridge length $r_l$
to be the distance between the points at which the red (or violet)
shift from a body at rest is half its maximum value. In the
spherically symmetric case, the ratios $r_0 /r_g$ and $r_l /r_g$
depend only on the equation of state of the bridge. Figure 3
presents the dependences of the ratios $r_0 /r_g$ and $r_l /r_g$
on $w =w_\paral= -w_\perp$ obtained numerically (see Appendix
8.1). The first diagram in this figure clearly shows why bridges
exist only when $w<-1$: the function $r_0 /r_g$ tends to zero as
$w\to -1$. Therefore, the radius of the throat decreases to zero,
and passage to another part of the Universe is no longer possible.

\section{FORMATION AND EXISTENCE OF BRIDGES}

The question of the formation (and, in general, the existence) of
bridges remain open and two scenarios are possible.

1. The trivial scenario: bridge were formed during the birth of
our part of the Universe and still exist, preserving the global
topology (geometry) of the Universe. This is the only way to form
bridges connecting two different parts of our Universe (like a
door handle), since the modern geometrical theorems of general
relativity forbid the disruption of such a topology [7]. We cannot
currently answer the question of how this topology was created
since Einsteinian gravitational theory cannot be applied to the
birth of the Universe (quantum gravity which remains incompletely
developed must be used).

2. The dynamic scenario: bridges could be formed instead of black
holes as "branches "in the process of evolution of the Universe.
This process of detachment is drawn schematically in Fig.4. First,
the matter (accumulated due to self-gravity) curves space. Next,
the amount of matter becomes sufficient to curve the geometry to
the limiting value at some radius (Fig.2). The matter then begins
to inflate the space behind this radius, similar to the inflation
of space by the vacuum condensate in the observed expanding
Universe (as a consequence of negative pressure). There must be no
event horizons anywhere. The resulting branch evolves by itself,
expands, and forms a new part of the Universe, connected to our
part of the Universe by the throat of the bridge.

\begin{figure}[t]
\centering \epsfbox[100 280 500 530]{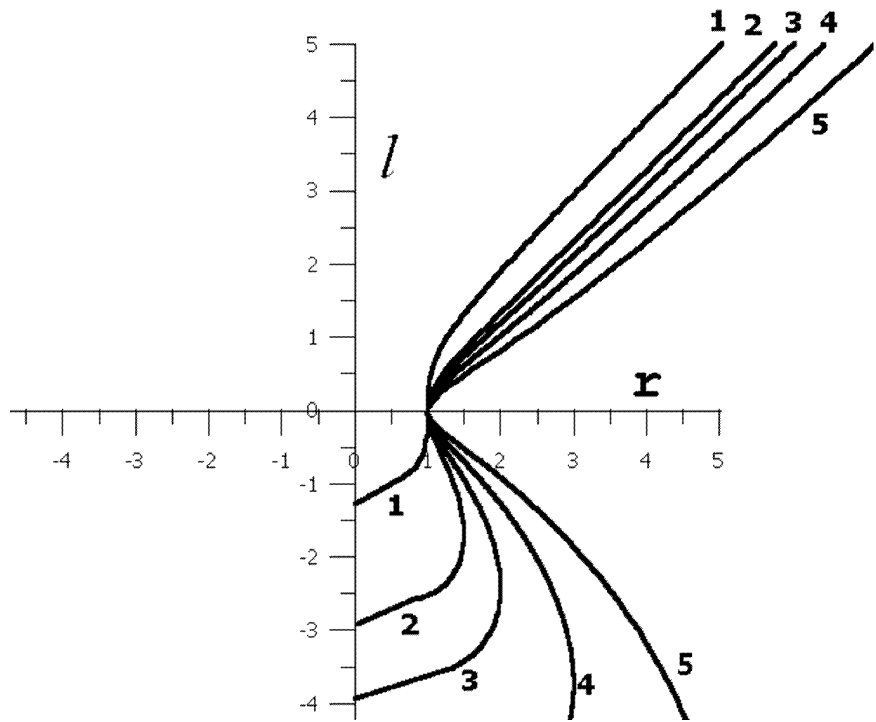}
 \caption{Probable dependence of $r$ on $l$ in a bridge during its
detachment. The numbers 1 to 5 correspond to successive times in
the bridge's evolution.}
 \label{R4}
 \end{figure}

\section{OBSERVATION OF BRIDGES}

The simplest way to detect a bridge is to observe some objects
(e.g. stars) through its throat. If the geometry of the Universe
is different on the opposite sides of the throat (see Appendix
8.2), the rate of flow of time will also be different. As a
result, photons passing the bridge's throat will experience a red
(or violet) shift. An important problem with this method is to
discriminate such shifts from Doppler shifts due to the source
motion.

Another method is based on the idea that the photons in light
signals passing through the throat of a bridge should experience a
deflection; i.e., they should undergo lensing by the curved
geometry of the space (see, for example,[8]). Let us define the
impact parameter $h$ for a photon deflected by a bridge to be $h
=L/E$ the ratio of the angular momentum and energy of the photon
at infinity.

There are two type of lensing (see Appendix 8.3).

1. The photon does not pass through the throat, and so always
remains in the same part of the Universe. This situation takes
place for photons that have sufficiently large impact parameters
(in comparison with the throat radius) and are deflected by the
gravitational lens.

2. The photon passes through the throat to another part of the
Universe. This situation takes place for photons having
sufficiently small impact parameters (compared to the throat
radius). One characteristic feature of this effect is that the
deflection of the photons is given by the formula for a thin
converging lens whose focal distance is approximately equal to the
throat radius. In the case of other gravitating centers, these
photons would be absorbed by the surface of the body.

\begin{figure}[t]
\centering \epsfbox[50 350 500 550]{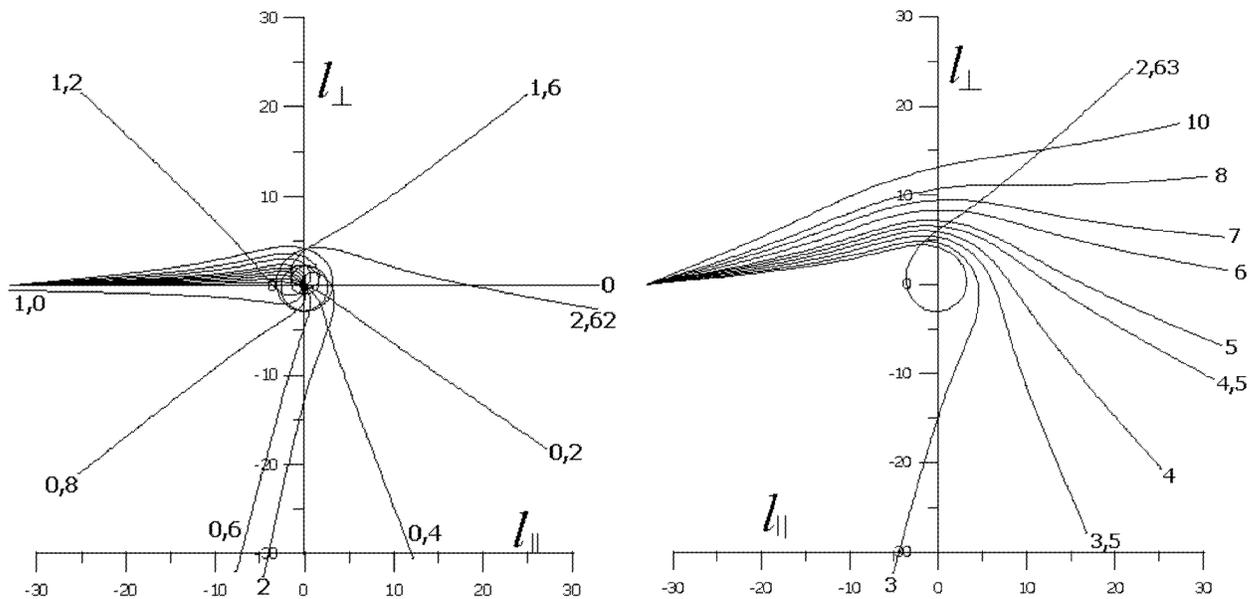}
 \caption{Deflection of photons by a bridge with $w = -2$. The trajectories of
photons deflected through an angle $\Delta\varphi$ from the
initial direction are plotted. The trajectories start at $l = -32
r_0$, $\varphi =-\pi$. Trajectories passing through the throat at
small impact parameters ($h<h_{cr}\approx 2.621 r_0$) and
terminating at $l\to +\infty$ are shown in the left-hand panel.
Trajectories not passing through the throat ($h>h_{cr}$) and
terminating at $l\to -\infty$ are shown in the right-hand panel.
The numbers near the ends of the curves indicate the impact
parameters (in units of $r_0$). The coordinates $l_\paral$,
$l_\perp$ denote the equatorial plane of the photon scattering
($l_\paral = +|l| \cos\varphi$, $l_\perp = -|l| \sin\varphi$). }
 \label{R5}
 \end{figure}

The calculated photon trajectories for a bridge with $w = -2$ are
presented in Fig.5. If the impact parameter is close to the
critical value $h_{cr}$, the photons can be deflected through
large angles (greater than $2\pi$) by the bridge. Note that
photons passing through the throat inevitably intersect the center
of the bridge; i.e., the point $l =0$. There is no singularity at
this point, since the value $l =0$ corresponds to a sphere of
finite radius $r_0$.

Since the photon impact parameter does not change and corresponds
to a location in the bridge 's throat (relative to its center)
from which we can observe the light of a star from another part of
the Universe, the impact parameter can be used to determine the
deflection angle of the photon on its way to the observer. In fact
this is a method to determine the coordinates of stars in that
part of the Universe.

\section{DISCUSSION}

Not only photons, but also physical bodies can pass through a
bridge. The main obstacles to this are tidal forces, which disrupt
bodies inside black holes and in the orbits of compact stars (such
as neutron stars). The tidal forces depend on the curvature of
space and the size of the body. Therefore, they coincide to order
of magnitude with the corresponding forces in black holes of the
same mass. The tidal forces affecting a body of mass $\mu$ and
size $d$ are proportional to the derivative of the gravitational
forces. For black holes, the tidal forces are equal to $\mu c^2
d/r^2_g$ at the event horizon. To order of magnitude, the same
forces should act in bridges. If a bridge possesses a sufficiently
large mass, then not only macroscopic objects, but also stars,
such as the Sun can pass through its throat without disruption.
This should be the case for bridges with masses of a few billion
solar masses (as in some quasars). A person would not feel the
passage through such a bridge, because the tidal forces would be
extremely small.

\section{CONCLUSIONS}

Our main conclusion is that bridges can be real objects, which are
described by Einstein gravitational theory in a self-consistent
way and can be distinguished from other celestial bodies by
observations. The equation of state of the matter in traversable
bridges described by general relativity must be substantially hard
and anisotropic [see (8)]. Gravitational lensing by bridges is
fundamentally different from lensing by other bodies, and
precisely these differences can be used to identify bridges and
study their properties (as well as the properties of the part of
the Universe on the opposite side of the bridge).
$$ $$

{\bf ACKNOWLEDGEMENTS

I am grateful to I.D. Novikov and N.S. Kardashev for their help
and participation in my preparation of this article. This work was
supported by the Program "Non stationary Phenomena in Astronomy",
the Program of Support for Leading Scientific Schools (project N
Sh-1653-2003.2), and the Russian Foundation for Basic Research
(project codes 01-02-16812, 01-02-17829, and 00-15-96698).}

\section{Appendices}

\subsection{NUMERICAL SOLUTION}

The Einstein equations corresponding to (6) and (7) are [9]:
\begin{equation}
2\varepsilon r^2 = - 2 r'' r + (1 - {r'}^2) ,
\label{1-1}\end{equation}
\begin{equation}
2\wpar\varepsilon r^2 = 2 r \phi' r' - (1 - {r'}^2) ,
\label{1-2}\end{equation}
\begin{equation}
2\wper\varepsilon = \phi'' + {\phi'}^2 + (\phi' r' + r'')/r
\label{1-3-2}\end{equation} These equations lead to the following
expression (which can be derived most easily from the equality
$T^k_{l;k}\equiv 0$:
\begin{equation}
\wpar (\ln\varepsilon)' = (\wper - \wpar)(\ln r^2)' - (1 + \wpar
)\phi' \, . \label{1-3}\end{equation} Here, a prime denotes a
derivative with respect to $l$ which is related to $r$ as
\begin{equation}
dl = \pm\sqrt{\exp (\la)}\, dr \, ,\quad l(r) =
\pm\int\limits_{r_0}^r \sqrt{\exp (\la)} \, dr \, ,\quad r(l) =
r_0 \pm \int\limits_0^l \sqrt{\exp (-\la)} \, dl \, .
\label{1-4}\end{equation} Here and below the upper and lower signs
correspond to positive and negative values of $l$ respectively.
The new constant $r_0$ appearing here is defined as the minimum of
the function $r(l)$: $r(0)\equiv 0$. Let us introduce the new
functions
\begin{equation}
a(l) \equiv r(1 - e^\nu) \, , \quad b(l) \equiv
r_0+\int\limits_0^l 2\varepsilon r^2 r'\, dl \, ,
\label{1-5}\end{equation} in which case (10) can be rewritten
\begin{equation}
b' \equiv 2\varepsilon r^2 r' = \left[ r(1-{r'}^2) \right]'
\label{1-6}\end{equation} The definitions (14) can be rewritten
\begin{equation}
(r')^2 = \exp (-\la) = 1 - b/r \, ,\quad l =
\pm\int\limits_{r_0}^r \left(1-b/r\right)^{-1/2}\, dr \, .
\label{1-7}\end{equation} Consequently the metric (6) will take
the form
\begin{equation}
ds^2 = (1 - a/r)\, dt^2 -  (1 - b/r)^{-1}\, dr^2 - r^2\, d\Omega^2
\, . \label{1-5-2}\end{equation} Equation (11) can be rewritten
\begin{equation}
\wpar b' \equiv 2\wpar\varepsilon r^2 r'= 2 (r-b) \phi' -r'b/r \,
. \label{1-8}\end{equation} In order for the quantities $a$ and
$b$ (corresponding to twice the mass of the system) to be limited,
the energy density $\varepsilon (r)$ must tend to zero at infinity
faster than $const /r^3$. The left-hand side of (19) then tends to
zero as $r\to\infty$ faster than the right-hand side, and can,
therefore, be neglected. In this limit, $\varphi'\to a r'/(2r^2)$.
Consequently $a$ and $b$ tend to the same limit, which is twice
the mass of the system or its gravitational radius:
\begin{equation}
\lim_{r\to\infty} a = \lim_{r\to\infty} b = 2m = r_g \, .
\label{1-8-1}\end{equation}

Since $b(l=0)=b_0 =r_0$ and $\lim_{r\to r_0}(r')\propto\sqrt{r-b}$
Eq. (19) gives
\begin{equation}
\varepsilon_0 = -{1\over 2\wpar r_0^2} \, .
\label{1-8-2}\end{equation} Further we can exclude the variable
$l$ from (13), (15), and (19) by substituting $r$. Using the new
variables
\begin{equation}
x=r/r_0\, ,\quad y=b/b_0\, ,\quad \xi =
\varepsilon/\varepsilon_0\, , \label{1-8-3}\end{equation} the
resulting equations can be reduced to the integral form
\begin{equation}
y(x)=1 - {1\over\wpar}\int\limits_{1}^x \xi x^2 \, dx \,  ,
\label{1-9}\end{equation}
\begin{equation}
\xi'(x) = \xi\left[({\wper\over\wpar}-1){2\over x} + {1\over
2}({1\over\wpar}+1){\xi x - y/x^2 \over 1 - y/x}\right]\, ,
\label{1-10}\end{equation}
\begin{equation}
\xi(x) =1 + \int\limits_{1}^x \xi' \, dx \, ,
\label{1-11}\end{equation}
\begin{equation}
\exp (\phi) = \exp (\phi_0)\cdot \xi^{-{1\over 1/\wpar +1}}\cdot
x^{2{\wper / \wpar - 1\over 1 / \wpar + 1}} \, .
\label{1-11-2}\end{equation} The factor $\exp (\varphi_0)$ can be
derived from the equality $\exp (\varphi (+\infty ))=1$. These
equations are closed and very convenient for numerical
integration.

The parameters $w_\paral$ and $w_\perp$ must satisfy two
requirements. The first was already noted before: the restriction
on the total mass of the system. This condition leads to the
inequality $x\xi' /\xi <3$ and, as can easily be derived from
(24),
\begin{equation}
2\wper/\wpar < -1 \label{1-12}\end{equation} The second
requirement is associated with the integrability of Eq. (17) for
$l$: the absence of an event horizon in the system. Since $y(1)=1$
the inequality $[y(x )/x ]'\le 0$. must be satis .ed at the point
$x =1$. This leads to the second condition \footnote{In general
the exact equality also corresponds to a static solution, but, in
this case, it follows from (19) that this solution possesses an
event horizon, $2\varphi =\ln (1-b/r )$; i.e., can not represent a
bridge.}:
\begin{equation}
1/\wpar < -1 \label{1-13}\end{equation} Conditions (27) and (28)
taken together lead to the two possible cases for the range of
$w_\paral$ and $w_\perp$.

The first case is physical, and is represented by inequality (8).
The second case is
\begin{equation}
\varepsilon < 0\, , \qquad 0 < \wpar < -2\wper \, .
\label{1-15}\end{equation} This case is of less practical
interest, since it supposes the existence of matter with negative
energy density. However we were able to obtain an analytic
solution for a bridge when $w_\paral = -w_\perp =1$ (see below).

It has not yet been possible to find an analytical solution for a
bridge in the first case, but solutions have been obtained
numerically (for example, for $w_\paral = -w_\perp = -2$ Fig.6).
This case is of the most interest, since a positive energy density
corresponds to positive masses. (It is quite unclear what negative
mass is, and to what type of matter it corresponds.)

\begin{figure}[t]
\centering \epsfbox[80 290 500 500]{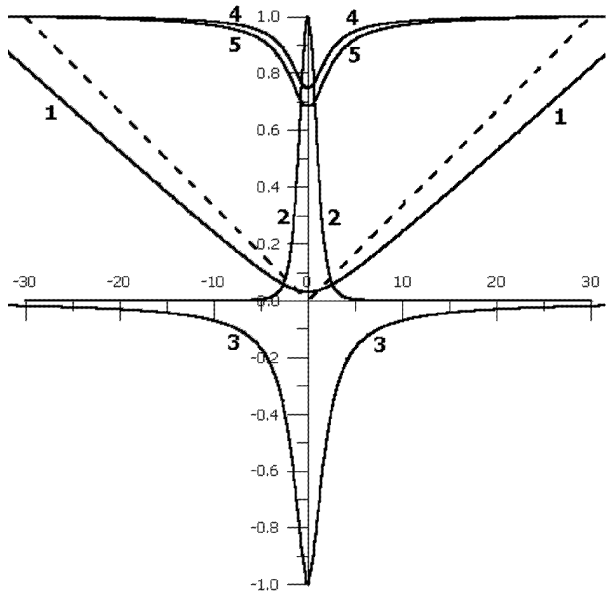}
 \caption{Numerical solution for a bridge with
$w_\paral = -w_\perp = -2$, $l_{max} = 30 r_0$: (1)
$r(l)/l_{max}$, (2) $\xi (l) = \varepsilon  (l)/ |\varepsilon_0|$,
(3) $\varphi (l)/ |\varphi_{max}|$, (4) $b(l)/r_g$, and (5)
$a(l)/r_g$. The lines $|l|/l_{max}$ are shown dashed.}
 \label{R6}
 \end{figure}

Equations (23)-(25) can be transformed from integral to
differential form. This is the most convenient form when $w
=w_\paral = -w_\perp$. Equations (23)-(25) are transformed to the
single nonlinear equation for $y(x)$
\begin{equation}
2(x-y)(xy''+2y') + (1+w)(x{y'}^2 + yy'/w) = 0 \, .
\label{1-11-4}\end{equation} The solution of this equation that is
finite at infinity can be written as the series
\begin{equation}
y(x)=\sum\limits_{n=0}^\infty C_n/x^n \, , \quad
\sum\limits_{n=0}^\infty C_n = 1 \, . \label{1-11-5}\end{equation}
This expression yields the following three exact solutions for the
static equations of general relativity.

1. The Schwarzschild solution for a black hole: $w$ is arbitrary
because
\begin{equation}
w \makebox{ is arbitrary, because }\varepsilon =0 \, , \quad y=1
\quad \makebox{ or } \quad b=r_g=r_0 \, .
\label{1-11-6}\end{equation}

2. The Reissner-Nordstrom solution for an electrically charged
black hole [10,11]:
\begin{equation}
w=-1\, ,\quad y=C_0 + C_1/x \quad \makebox{ or } \quad b = r_g -
Q^2/r \, . \label{1-11-8}\end{equation} where $Q$ is the electric
charge.

These solutions can easily be verified using formulas (17)-(19).

3. The exact analytic solution for a bridge in the case $w =1$
takes the form
\begin{equation}
w=+1\, ,\quad y_{w=1}=1/x \quad \makebox{ or } \quad b = r^2_0/r
\, . \label{1-11-7}\end{equation} This solution is described in
detail in the next Appendix.

\subsection{ANALYTIC SOLUTION}

The metric of the well-known solution for a stable macroscopic
bridge [6] is ($w =w_\paral = -w_\perp =+1$)
\begin{equation}
ds^2=(1-2\eta/R)^{m/\eta}\, dt^2 - (1-2\eta/R)^{-m/\eta}\, dR^2
-(1-2\eta/R)^{1-m/\eta}\, R^2\, d\Omega^2 .
\label{5-3}\end{equation} Here, $\eta^2 =m^2 +q^2$ are the mass
and charge of the bridge. The metric (35) reduces to the
Schwarzschild metric if the charge is zero.

By substituting for the coordinate $R$ this metric can be reduced
to real form in the entire space for
\begin{equation}
ds^2 = \exp (-2m/\tilde R)\, dt^2 - \exp (2m/\tilde R) \left[ dR^2
+ (|\eta|^2 + R^2)\, d\Omega^2 \right] \label{5-3-2}\end{equation}
Here, $\cos^2 ( |\eta  |/\tilde R)= R^2 /( |\eta |^2 +R^2 )$.

Using the coordinate transformation $r^2 =(|\eta |^2 + R^2) \exp
(2m/\tilde R)$, the metric can be expressed in terms of the
coordinate $r$:
\begin{equation}
ds^2 = {|\eta|^2 + R^2\over r^2}\, dt^2 - {|\eta|^2 + R^2\over (R
- m)^2}\, dr^2 - r^2\, d\Omega^2 \label{5-4}\end{equation} The
corresponding derivative takes the form
\begin{equation}
{dr\over dR} = {(r/|\eta|)\cdot (R/|\eta| - m/|\eta|)\over
1+R^2/|\eta|^2}\, . \label{5-4-1}\end{equation} We can now find a
relation between $R$ and $l$:
\begin{equation}
l = R - m \quad\makebox{ at }\quad r = r_{min}\, .
\label{5-4-4}\end{equation} The corresponding right-hand sides of
the Einstein equations take the form
\begin{equation}
2\varepsilon = 2 P_{\parallel} = -2 P_{\perp} = {q^2\over r^2\cdot
(R^2 - \eta^2 )} < 0 \, . \label{5-4-2}\end{equation} If $\eta$ is
imaginary the coordinate $r$ has a minimum, to the bridge's
throat. The bridge is completely defined by the quantities
\begin{equation}
\begin{array}{lll}
r = |\eta|\cdot \sqrt{1+R^2/|\eta|^2} \cdot
\exp\left\{(m/|\eta|)\left[\pi/2-\atg(R/|\eta|)\right]\right\} \, ,\\
\phi = (-m/|\eta|)\left[\pi/2 -
\atg (R/|\eta|)\right] \, ,\\
b/r = \left( 1 + 2 m R / |\eta|^2 - m^2/|\eta^2|\right) /
(1 + R^2/|\eta|^2) \, ,\\
V^2 = 1 - \exp\left\{(2m/|\eta|)\left[ \atg
(R/|\eta|)-\atg(R_0/|\eta|) \right]\right\} \, .
\end{array}
\label{5-9}\end{equation} We can see that this solution is
symmetric about the throat only in the case of zero mass m
(Fig.7). In this case, the expression for $b$ coincides with (34).

\subsection{DEFLECTION OF PHOTONS
BY THE FIELD OF THE BRIDGE}

To calculate a photon trajectory in the gravitational field of a
bridge, we use the theory described in [9,$\S$ 101]. A photon
eikonal in the equatorial plane of a centrally symmetric field
will take the form
\begin{equation}
\Psi = -Et + L\varphi + \Psi_l, \label{6-1}\end{equation} where
$E$ and $L$ are the conserved energy and angular momentum of the
photon, $\Psi_l (l)$ is the radial part of the eikonal, and
$\varphi$ is the azimuthal angle in the equatorial plane. The
Hamilton-Jacobi equation for this eikonal is
\begin{equation}
e^{-\nu}(\partial_t \Psi)^2 - (\partial_l \Psi)^2 -
(\partial_\varphi \Psi)^2/r^2 = 0 \, . \label{6-2}\end{equation}
We thus obtain:
\begin{equation}
\Psi_l = \int \sqrt{E^2/(1-a/r) - L^2/r^2} \, dl \, ,
\label{6-3}\end{equation}
\begin{equation}
t-t_0 = \int\limits_{l_0}^l {E/(1-a/r)\, dl\over \sqrt{E^2/(1-a/r)
- L^2/r^2}} \, , \label{6-4}\end{equation}
\begin{equation}
\varphi -\varphi_0 = \int\limits_{l_0}^l {(L/r^2)\, dl\over
\sqrt{E^2/(1-a/r) - L^2/r^2}} \, , \label{6-5}\end{equation} Let
the photons be emitted in that part of the Universe characterized
by negative values of $l$. Then, these photons enter our part of
the Universe (with positive values of $l$) when they pass through
the bridge. We obtain from the previous three expressions
\begin{equation}
\begin{array}{lll}
\partial_t l = \pm \sqrt{(1-a/r) - (1-a/r)^{2} h^2/r^2} \, ,\\
\partial_t \varphi = h (1-a/r)/r^2 \, ,\\
\partial_t r = \mp (1-a/r)\sqrt{(1-b/r)((1-a/r)^{-1} - h^2/r^2)} \, .
\end{array}
\label{6-6}\end{equation}

\begin{figure}[t]
\centering \epsfbox[100 280 500 490]{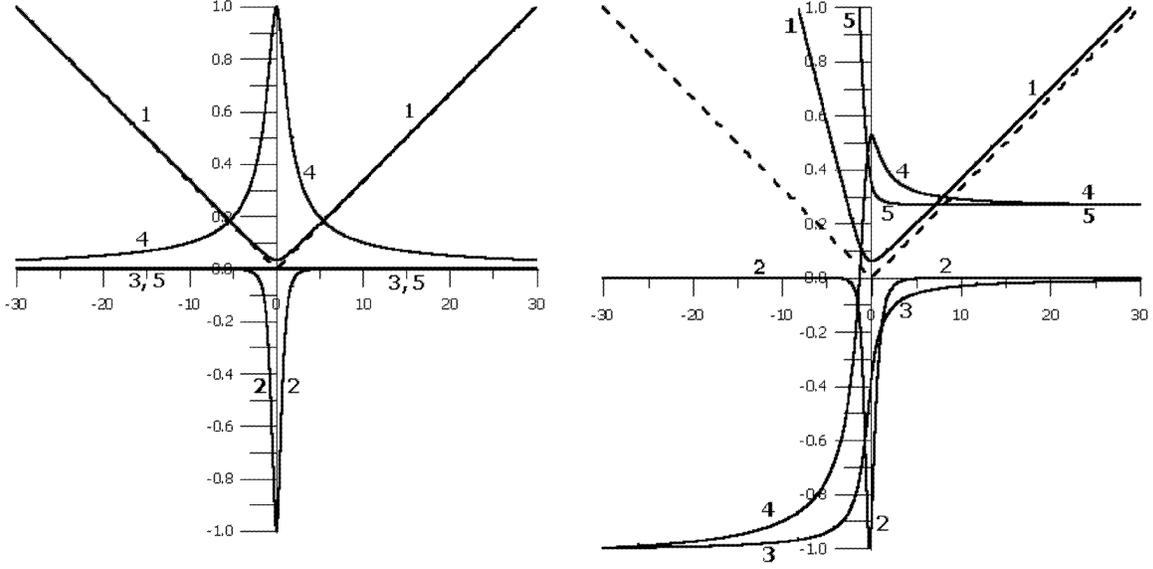}
 \caption{Analytic solution for a bridge with $w_\paral = -w_\perp = +1$ at
$l_{max} =30 r_0$: (1) $r(l)/l_{max}$, (2) $\varepsilon
(l)/|\varepsilon_{max}|$, (3) $\varphi (l)/ |\varphi_{max}|$, (4)
$b(l)/ |b_{max}| \to r_g / |b_{max}|$, and (5) $a(l)\cdot  const
\to r_g / |b_{max}|$. The lines $|l|/l_{max}$ are shown dashed.
The left-hand panel represents the symmetric solution for $m=0$
and the right-hand panel the asymmetric solution for $m=0.2|\eta
|$.}
 \label{R7}
 \end{figure}

Particular attention should be paid here to the signs $\pm$. The
equation for the trajectory $\varphi (r)$ is obtained by
integrating over the radius $r$. The sign of the integrand changes
together with the changing sign in (47) when the point of minimum
radius is passed. In the limiting case of infinite initial and
final radii, the change of the angle $\varphi$ is
\begin{equation}
\Delta\varphi = 2\int\limits_{r_{min}}^{\infty}\, {h\over
r^2\sqrt{(1-b/r)((1-a/r)^{-1} - h^2/r^2)}} \, dr \, .
\label{6-7}\end{equation} This expression can be integrated in
explicit form in two important limiting cases:

(1) $h\ll r_0$, $h\ll a$, $h\ll b$ is the case of when the impact
parameter is sufficiently small that the photon can pass through
the throat. In this case, $r_{min} =r_0$ and the $h^2 /r^2$ term
in the denominator of (48) can be neglected. Assuming that $a$ and
$b$ are constant, we obtain
\begin{equation}
\Delta\varphi_{_I} = {2h\over b} \left\{ 1-\sqrt{({b\over
r_0}-1)({a\over r_0}-1)} + {b-a\over 2\sqrt{ab}} \ln \left[
{(\sqrt{b(a-r_0)}+\sqrt{a(b-r_0)}(\sqrt{b}-\sqrt{a}) \over
(\sqrt{b(a-r_0)}-\sqrt{a(b-r_0)})(\sqrt{b}+\sqrt{a}) }\right]
\right\} \, . \label{6-8}\end{equation} This formula is analogous
to that for a converging thin lens.

(2) $h\gg r_0$, $h\gg a$, $h\gg b$ is the case when the impact
parameter is sufficiently large that the photon does not pass
through the throat. In this case, $a\approx b\approx r_g$ is a
constant and $r_{min}$ is determined by the equation $r^2_{min} +
a r_{min} -h^2 =0$. Neglecting terms quadratic in $a/h$ and $b/h$
and substituting the variable $z =r -b/2$ we obtain
$z_{min}\approx h -(a + b)/2$,
\begin{equation}
\Delta\varphi_{_{II}} = \pi + {a+b\over h} \, .
\label{6-9}\end{equation} This expression coincides with the usual
formula for the gravitational lensing of photons when $a =b =r_g$.

When a photon trajectory corresponds to the first or second cases
depends on the value of the impact parameter compared to the
critical value defined by the equation
\begin{equation}
re^{-\phi}=h_{cr}=min \quad \makebox{ or }\quad
1-r_{cr}\phi'(r_{cr})=0\, . \label{6-6-0}\end{equation} For
example for a bridge with $w = -2$, one can find numerically that
$h_{cr}\approx 2 .621$.

{\bf Translated by Yu.Dumin}

\end{document}